\documentstyle[twocolumn,aps,epsfig]{revtex}
\begin{document}
\newcommand{\beq}{\begin{equation}}
\newcommand{\bef}{\begin{figure}}
\newcommand{\eeq}{\end{equation}}
\newcommand{\efi}{\end{figure}}

\newcommand{\bfxi}{\mbox{\boldmath$\xi$\unboldmath}}

\twocolumn[\hsize\textwidth\columnwidth\hsize\csname @twocolumnfalse\endcsname
\title{A Real-Space Full Multigrid study 
       of the fragmentation of ${\rm Li}_{11}^{+}$ clusters.}
\author{Francesco Ancilotto, Philippe Blandin and Flavio Toigo}
\address{ INFM-Dipartimento di Fisica "G.Galilei"- Universita'
          di Padova, via Marzolo 8, I-35131 Padova, Italy}
\maketitle

\begin{abstract}
We have studied
the fragmentation of ${\rm Li}_{11}^+$ clusters into the two experimentally 
observed products ${\rm Li}_9^+ + {\rm Li}_2$ and ${\rm Li}_{10}^+ + {\rm Li}$. 
The ground state structures for the two fragmentation channels are found by 
Molecular Dynamics Simulated Annealing in the framework of Local 
Density Functional theory. Energetics considerations suggest that 
the fragmentation process is dominated by non-equilibrium processes.
We use a real-space approach to solve the Kohn-Sham problem, where
the Laplacian operator is discretized according
to the {\it Mehrstellen} scheme, and take advantage
of a Full MultiGrid (FMG) strategy to accelerate convergence. 
When applied to isolated clusters,
we find our FMG method to be more efficient than state-of-the-art plane wave 
calculations.
\bgroup\draft
\pacs{PACS numbers:66.10Ed,71.55Jv,72.20Jv}\egroup
\end{abstract}
]

\noindent
\section{Introduction}
In recent years a number of real-space grid-based algorithms have been proposed
to study the electronic properties of condensed matter systems.
Most of these approaches, based on Density Functional (DF) theory,
evaluate in real space the action of the effective potential and of
the kinetic energy operators that enter the one-electron DF equations.

When based on an accurate approximation of the Laplacian operator,
real-space methods 
exhibit several advantages when compared to Plane Waves (PW) schemes. 
In particular:

a) Boundary conditions are not constrained to be periodic,
   thus facilitating the study of charged 
   systems or of systems having finite electric multipole moments (e.g. isolated
   clusters). 

b) In real space, it is technically easier to increase the resolution 
   of the grid locally. 
   This feature is 
   particularly appealing in the case of first-row and transition-metal 
   elements and for the
   study of clusters: a higher density of grid points is usually needed in 
   the regions where the ions are located, whereas a lower density is 
   sufficient on the vacuum side to describe 
   the exponentially small tails of the electronic wavefunctions. 
   The use of locally refined 
   meshes may lead to a significant reduction of the computational effort.

c) With a suitable discretization of the Laplacian operator, all the 
   operations needed to
   compute the total energy of the system are short ranged, hence 
   facilitating the parallelization of the computer code.

Among the various approaches reported in the literature, let us mention the 
use of high-order finite 
difference methods for representing the laplacian, combined with 
the use of soft non-local pseudopotentials on uniform grids  \cite{cheli} to 
calculate the electronic structure and the short time dynamics of small 
molecular systems. 
Gygi et al. \cite{galli} extended the real-space adaptive-coordinate 
method of Ref.
\cite{gygi} to perform LDA electronic structure calculations of small 
molecules. 
Ref \cite{zumbach} reports a similar approach. Other recent real-space 
methods include the 
use of wavelet basis sets \cite{cho,wei}, the finite-elements method 
\cite{white,tsu} and the 
Multigrid (MG) calculations of first and second row atoms \cite{iyer}. 

Briggs et al. \cite{briggs,brig96} have recently proposed 
an efficient method that combines an accurate discretization of the 
Laplacian ({\it Mehrstellen})
with a MG acceleration scheme to solve
Poisson and Kohn-Sham (KS) equations in large-scale LDA calculations. 

Our real-space calculations follows 
Briggs' et al. \cite{briggs,brig96}
in the use of the ({\it Mehrstellen}) discretization scheme,
but differs from previous calculations in a significant way in the 
scheme chosen to accelerate convergence.
Whereas in Ref.\cite{brig96} a MG solver was used to
solve Poisson and KS equations we implement here a Full MultiGrid (FMG) 
diagonalization procedure.

Given some similarities between our method and the one 
of Ref.\cite{brig96}, we will 
concentrate on their differences and refer the interested reader 
to Ref.\cite{briggs,brig96} 
for technical details common to both approaches.

This paper is organized as follows: in Section II we discuss our 
Full Multigrid (FMG) algorithm, in Section III we demonstrate its accuracy 
and performances by running a few tests on simple molecules. 
We then apply our scheme to the study 
of the fragmentation of 
${\rm Li}_{11}^+$ clusters in Section IV. Section V contains a summary and 
some concluding remarks.

\noindent
\section{Computational method}

\subsection{Multigrid Methods}

MG methods were introduced in the 70's by Brandt \cite{mg} as a tool to 
solve elliptic PDE discretized on a 
grid of $N$ 
points in $O(N)$ operations. His idea was to use some auxiliary set of grids 
to speed up the convergence of straight relaxation methods \cite{shaw}. 
In these methods
the value at the mesh points are
repeatedly updated according to the discretized differential equation.
This process enforces a local consistency between the updated value 
and that of its neighbors. By performing enough relaxation
cycles the exact solution propagates from the boundaries, where it is fixed
by the boundary conditions, to the interior of the grid.
Relaxation
iterations reduce quickly the components of the error with wavelength comparable
to the grid spacing but remain quite ineffective in reducing error components
with larger wavelength.
The MG strategy aims at removing these low frequency errors by noticing that
a) low frequencies on a fine grid become higher frequencies on a coarser grid,
and b) the error on the solution of a PDE with fixed boundary conditions
satisfies a similar PDE, with zero boundary conditions. Solving this PDE for
the error on a coarse grid allows to remove the low
frequency errors that spoil the fine grid solution. This two-level
strategy, often referred to as ``V-cycle''\cite{hackbush,recipes}, 
can naturally be applied recursively and involve an arbitrary number of levels.
These ``V-cycles'' allow to converge very rapidly to the right solution. 
They enhance drastically the efficiency of relaxation methods and are referred 
to as {\it the} MG method.
The interested reader is referred to Ref.
\cite{hackbush} and to references therein for more details. 

MG methods are particularly well adapted to solve Poisson equation and to
large scale eigenvalue problems \cite{mg,costiner}.

An improvement over MG methods is made possible by using a FMG strategy.
While both approaches make use of Brandt's idea and use the auxiliary set 
of grids to perform ``V-cycles'', the FMG approach is 
somewhat more complete in that it also uses these grids to construct a 
good first guess at almost no cost. In other words, while MG codes start 
the computation on the final grid and resort to ``V-cycles'' to speed up 
convergence, FMG codes solve the problem on some coarse grid first and,
once the solution is converged well enough, expand it on the next finer grid.
The coarse grid solution is a preconditioning for the initial guess on 
the next finer grid. This preconditioner is used up to the finest grid.
FMG codes take obviously advantage of the ``V-cycles'' on each of the 
intermediate levels.

As a consequence,
FMG schemes further enhance the 
rate of convergence of MG algorithms since they reduce the residuals at each 
iteration \cite{hackbush}. In practice, FMG is only slightly more complex 
to implement than MG.

\subsection{Grid representation of the Kohn-Sham problem.}

In the Kohn-Sham \cite{ks} formulation of Density Functional Theory \cite{hk},
the total energy $E[\{ \psi_i\},\{\bf{R}_{a}\}]$ of a system of $N_{a}$ ion 
cores 
located at $\{{\bf R}_{a}\}$ and of $N$ electrons writes (in atomic units and 
considering only doubly occupied states):
\begin{eqnarray}
E[\{ \psi_i \},\{{\bf R}_{a}\}] &=& 
      2 \sum_{i=1} ^{occ} \int \psi_i^\ast [-{1 \over 2}\nabla
            ^2 ] \psi_i d{\bf r} \nonumber \\
        &+&{1 \over 2}\int \int {\rho ({\bf r}) \rho ({\bf r}^ \prime )  \over
|{\bf r}-{\bf r}^ \prime|  } d{\bf r}d{\bf r}\prime \nonumber\\ 
&+& 
\int ({\bf r}) V_{ion}({\bf r}) \rho ({\bf r})  d{\bf r}
\nonumber \\
&+& E_{xc} [ \rho ({\bf r})] + E_{ion,ion} (\{{\bf R}_a\})
\label{energy}
\end{eqnarray}
where the $\{ \psi_i \}$'s are the Kohn-Sham (KS) electronic orbitals.

The first, second and third terms of the right hand side are the kinetic, 
the electron-electron (Hartree) and the electron-ion energies, respectively. 
The fourth term stands 
for the exchange-correlation contribution, while the last one corresponds to the 
ion-ion electrostatic repulsion.
The electron density is given in terms of the $N/2$ 
occupied KS orbitals by:

\beq
\rho ({\bf r}) = 2 \sum_{i=1}^{occ} |\psi _i ({\bf r})|^2,
\label{rho}
\eeq
where the KS orbitals are the solutions of:
\begin{equation}
[-{1\over 2}\nabla ^2 + V_{eff} ]\psi _i({\bf r})=\epsilon _i \psi _i({\bf r}),
\protect\label{KS_eqn_1}
\end{equation}
subject to the orthonormality constraint $<\psi _i|\psi _j>=\delta _{ij}$. 

The effective potential $V_{eff}({\bf r})$ is given by:

\beq
 V_{eff}({\bf r}) = V_{ion}({\bf r}) + V_H({\bf r})+ V_{xc}({\bf r})
\label{veff}
\eeq
where 
$ 
V_H({\bf r})=\int \frac{\rho({\bf r^\prime})}{|{\bf r}-{\bf r}^\prime |} d{\bf r}\prime
$
is the Hartree potential and
$V_{xc}({\bf r})\equiv  \delta E_{xc}/\delta \rho({\bf r})$ 
the exchange-correlation potential. 
In our calculations the electron-ion interaction $V_{ion}$ 
is represented by {\it ab initio} 
norm-conserving pseudopotentials.
We use the Kleinman-Bylander \cite{kleinman} form in real space
\cite{martins} for the non-local part of $V_{ion}$.

For simplicity we reported the equations for paramagnetic systems i.e. for systems 
with all states doubly occupied. The generalization of Eqs. (\ref{energy}) and (\ref{veff}) 
to spin-polarized system (Local Spin Density, LSD) is straightforward \cite{vosko}.
For our LDA and LSD computations we resort to the Perdew and Zunger \cite{Perdew} 
parametrization of the Monte Carlo data of Ceperley and Alder \cite{Ceperley}.

Following Ref.\cite{briggs}, we discretized the Laplacian operator in
Eq. (3) according 
to a 
Mehrstellen scheme of order $O(h^{4})$ \cite{mehr}. The latter was indeed proven to provide at 
least the accuracy of schemes using a sixth-order central finite-difference 
discretization of the Laplacian \cite{briggs,brig96}, with the advantage of using
more local data (i.e. only up to second-nearest neighbors for a 
cubic mesh).

The use of a uniformly spaced grid with spacing $h$ 
allows to define an effective energy cut-off 
equal to that of the PW calculations that use the same real-space grid 
for their FFT's. Assuming that the electronic charge density in the plane wave code 
is expanded with an energy cut-off four times as large as that of the wavefunctions, 
we define $G_{max}^2=\pi ^2/4 h^2\,\,Ry$.

In our code all integrations are performed according to the three-dimensional 
trapezoidal rule:
\beq
\int g({\bf r})d{\bf r} = h^3\sum _{ijk} g({\bf r}(i,j,k)).
\label{integ}
\eeq
We notice that for high accuracy, it is essential that all the integrands
$g({\bf r})$ be band-limited in the sense that their Fourier transform must 
have zero magnitude in the frequency range $G>G_{max}=\pi/2h$. While this
condition is automatically fulfilled in PW calculations, since the basis set 
is cut-off at a specific plane-wave energy, in real-space calculations
high-frequency components above the natural cut-off $G_{max}^2$ can manifest 
themselves on the grid. In particular, if the pseudopotential $V_{ion}$ 
contains high frequency components near or above $G_{max}$, these high frequency 
components may be aliased to lower frequencies in an unpredictable manner 
\cite{briggs} leading to unphysical variations in the total energy as the ions 
move with respect to the grid.

This defect can be overcome by explicitly eliminating the high frequency components 
of the pseudopotentials by Fourier filtering, as shown in the context of PW 
by King et al. \cite{king}. In this work, we followed their prescription 
and removed from the non-local pseudopotential the Fourier components near $G_{max}$. 
This filtering procedure doesn't introduce any significant 
computational overload since it is done 
once for all at the beginning of the calculation. 
In real space every update of the non-local part of the total energy
and of ionic forces scales as the square of the number of atoms in the
unit cell, as opposed to the scaling as the cube of the system size
in the conventional reciprocal-space formulation \cite{king}.

\subsection{Full MultiGrid approach to the solution of the Kohn-Sham problem.}

In order to determine the ground-state electronic density for a given
configuration of ionic cores, we solve self-consistently  Eq. (\ref{KS_eqn_1})  
using a FMG approach along the lines described below.

In our calculations we use a cubic simulation cell of volume $L^{3}$ with 
$k_{max}$ uniform samplings. Calling $k=0,1,...k_{max}$ the level number, the grid 
at level $k$ contains $N^3_k$ points spaced by $h_k=L/(N_k-1)$, with $N_k=2^k(n_0-1)+1$. 
In most cases we use three grid levels $k=0,1,2$, refered to as ``coarse'', 
``intermediate'' and ``fine'' in the following. On the ``coarse'' and ``intermediate'' 
grids we take $V_{ion}$ as purely local and approximate it with the $s$-component of the 
non-local pseudopotential. The choice of the coarsest grid and thus of $n_0$, is dictated 
by the constraint that it must contain enough points to allow for a meaningful
representation of all the KS orbitals we are interested in: i.e. it must allow for the  
orthogonalization of the KS orbitals and still leave them enough degrees of freedom to 
build up a reasonable density.

On every grid, the relaxations are of the Gauss-Seidel type \cite{hackbush} and are 
performed according to the red-black ordering. The restriction and interpolation
procedures used in the MG-cycles are the projection of a weighted average and a simple 
tri-linear interpolation. The averaging procedure involves the first, second and third 
nearest neighbors of the fine point to be transferred.

Our FMG solver starts on the ``coarse'' level of $n_0^3$ points, where we perform
a few self-consistent iterations only. The KS orbitals are then transferred to the 
``intermediate'' grid. Once the orbitals are converged well enough they are 
interpolated on the ``fine'' grid. 
We solve on the latter 
the final KS problem. On the intermediate and fine grid we take 
full advantage of the MG
strategy we will describe in the following.

We may identify here a first fundamental difference between our approach and that 
of Briggs et al. \cite{brig96}: while they solve the KS problem directly on the fine 
grid and resort to the auxiliary set of coarse grids in a MG strategy, 
we, in contrast, implement a FMG solver in order to precondition the initial KS 
orbitals on the fine grid and thus start our computations on the coarsest grid. 
Possible advantages of our approach are discussed in Section III.
We will outline another important difference in subsection 2 below.

\subsubsection{Poisson equation}

The Hartree potential $V_{H}$ in Eq.\ (4) is the solution of the Poisson equation:
\begin{equation}
 \nabla^{2} V_{H} = -4 \pi \rho
\label{Poi_eqn}
\end{equation}
where $\rho$ is the electronic charge density. 

Since we are interested here in isolated clusters, we determine the boundary conditions 
on the potential $V_H $ from the multipoles of the charge density. When the simulation
cell is large enough, this is a good approximation that allows the study of charged as 
well as of neutral systems. Other boundary conditions (e.g. periodic) can be imposed 
without difficulty \cite{briggs}.

Our FMG Poisson solver does usually have more levels then the KS-solver. The coarsest grid 
for Poisson has $n_{0P}^3$ points and $n_{0P}$ relates to the $n_{0}$ of the KS problem 
by $n_{0}=2^j (n_{0P}-1)+1$ with $j$ such that $n_{0P}$ is the smallest 
possible integer larger than one. 
On the coarsest grid we use an exact solver, a direct matrix inversion if necessary.
The determination of $V_{H}$ on the set of coarse grids is of course possible since we
know $\rho$ on a fine grid. The ``V-cycles'' for Poisson are based on the following set of 
equations \cite{hackbush}: 
\beq
 \hat{V}_{H}=V_{H}+e
 \nonumber
\eeq
\beq
 r=4 \pi \rho+ \nabla^{2} \hat{V}_{H}
 \nonumber
\eeq
\beq
 \nabla^{2} e=r
\label{equ_pois_}
\eeq
$e$ stands for the error on the actual solution $\hat{V}_{H}$ and the residual $r$ is a 
measure of this error. We can determine $e$ by solving Eq.(9) for
zero boundary conditions.
FMG approaches allow to converge the solution of the Poisson equation
to the desired accuracy with a minimum number of V-cycles on the finest grid.
Notice that the addition of the extra coarse grid levels is computationally
inexpensive since at each level the number of grid points is reduced by
a factor of approximately eight.

\subsubsection{Kohn-Sham equations}
Solving the KS equations (\ref{KS_eqn_1})
is obviously far more complex than solving the Poisson problem since 
Eqs. (\ref{KS_eqn_1}) are non-linear: wavefunctions and eigenvalues must be determined
simultaneously.
In the spirit of the FMG procedure, we start our computations with some guess
on  the
``coarse'' grid, and get there a solution as accurate as possible.

Since we already explained the basics of our FMG approach to the KS problem,
we just state here the equations
our ``V-cycles'' are based on. 
In the following $\hat{\psi}_{i}$ stands for the actual solution, $\psi_{i}$ for
the exact solution, and $r$ for the residual associated to $\hat{\psi}_{i}$.
\beq
  \hat{\psi}_{i}=\alpha \psi_{i} + {\psi_{i}}_{\perp}  
  \protect\label{resid}
\eeq
\beq
   r=-\frac{\nabla^{2}}{2} \hat{\psi}_{i} + (V - \epsilon _{i})\hat{\psi}_{i}
\eeq
\beq
  -\frac{\nabla^{2}}{2} {\psi_{i}}_{\perp} + (V-\epsilon _{i}) {\psi_{i}}_{\perp} = r_{\perp}
   \protect\label{defect2}
\eeq
We determine an approximate ${\psi_{i}}_{\perp}$ by a few relaxions according to 
Eq. (\ref{defect2}) and use Eq.(\ref{resid}) to improve $\hat{\psi}_{i}$.

We evaluate the eigenvalue $\epsilon_{i}$ associated to $\{ \psi_i \}$ by resorting to 
Rayleigh's formula \cite{recipes} 
and ensure the orthogonality of the states with a Gram-Schmidt procedure.

In order to avoid instabilities due to charge ``sloshing'', we mix the new and old densities:
$\rho_{new}=(1-\alpha)\rho_{old}+2 \alpha \sum _{i=1}^{occ} \psi_i^2$, where 
$0 < \alpha \leq 1$ (usually a value of $\alpha $ between 0.3 and 0.5 is used in our 
calculations).

At this point we stress a second significant difference between our scheme and that of Briggs 
et al.\cite{brig96}. While we base our V-cycle on (\ref{defect2}) they resort to the Poisson 
equation: 
\begin{equation}
\nabla^{2}  \psi_{\perp} = r,
\protect\label{pois1}
\end{equation}
implicitly assuming that the leading terms entering the residual $r$ are due to an  
overestimate of the kinetic energy contribution. Furthermore, while we use the Mehrstellen
discretization on every grid level, they use it on their finest grid only. On the 
other grids they approximate the Laplacian operator with a 7-point central finite-difference.

To conclude this section, we note that, as suggested by Briggs et al. \cite{bernholc,brig96},
it is crucial to perform a Ritz projection \cite{hackbush} every now and then during the
self-consistent iterations in order to unmix eigenstates that may be close in energy. This
procedure improves the convergence to the exact self-consistent solution. 

\section{Tests and Results}

In order to check the accuracy and robustness of our approach and to compare its efficiency 
with conventional PW calculations we made a number of tests on small molecules.

First of all, we verify that the spurious effects 
due the components of the pseudopotential with frequencies higher than $G_{max}=\pi/2h$ are 
actually made very small when
the pseudopotentials
are filtered according to the procedure \cite {king} mentioned in Section 2.B.
These spurious effects usually appear as an  
unphysical dependence of the total energy  
on the positions of the ions relative to the  grid points.

To check this we consider an isolated $C_2$ molecule in the middle of a cubic cell and 
assume zero boundary conditions on the wavefunctions. We make a series of calculations where 
the $C_2$ dimer is displaced rigidly relative to the grid along a direction parallel to the 
cell edge and evaluate the total energy for each position.
The largest displacement from the center of the cell is $h/2$, where $h$ is the grid spacing
of the finest grid used in the calculation.
We assume a cubic cell of side $L=14\,{\rm a.u.}$, which we checked to be large enough to 
justify our zero boundary conditions, and use three uniform grids of $17^3$, $33^3$ 
and $65^3$ points. The Poisson equation is solved using additional coarser grids down 
to a mesh of $3^3$ points. The finest grid used for the KS problem corresponds
to a plane-wave energy cut-off $E_{cut}=52$ Ry, which is sufficient to give reasonably 
well converged properties of Carbon systems. We use {\it ab-initio} norm-conserving 
pseudopotentials \cite{bachelet} with s-non locality. 

Usually about 20 self-consistency cycles are sufficient to achieve a good convergence 
for a given dimer position. Every of these cycles consists of $2$ V-cycles, all the
relaxations are repeated twice and the eigenvalues are updated after every V-cycle.
We perform the Ritz projection, mentioned in the previous Section, every 5 
self-consistent cycle. 

We find that the variation of the total energy as the dimer is displaced across the cell 
is at most $0.5\,meV$,
showing that aliasing effects are indeed very small,  
at least on the scale of energies we are interested in.

We have also calculated the equilibrium distances and the vibrational frequencies of a number 
of simple molecules ($C_2, O_2, CO,Si_2$) and compared our results with experiments.
When possible, comparisons were also made with the results obtained with
a PW code using an energy cut-off corresponding to the mesh used in our real-space calculation.
The two sets of results are in very good agreement with each other since we didn't find 
discrepancies exceeding $1\%$ both for the equilibrium distances and for vibrational 
frequencies.

\begin{figure}[htb]
\centerline{\psfig{file=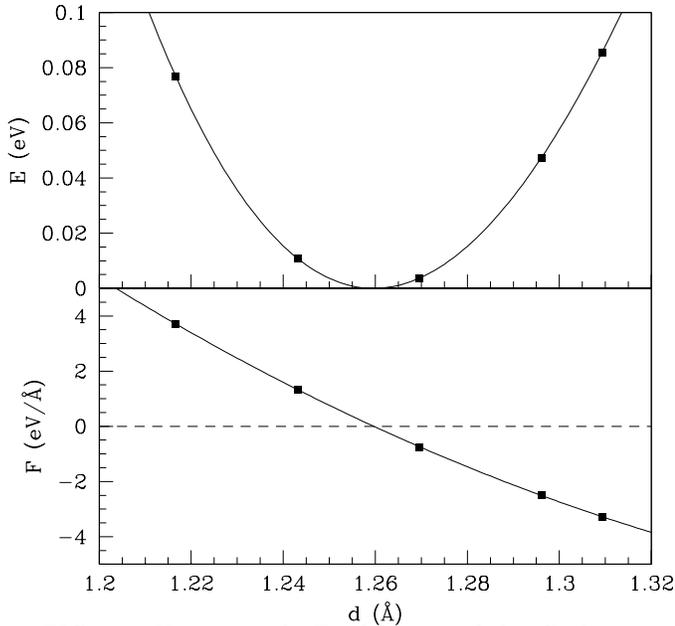, width=9cm, angle=0}}
\caption{ Upper panel: Total energy of the $C_2$ dimer as a function of the $C-C$ separation.
          The squares are the calculated points, the solid line is a fit with a $4-th$ order
          polynomial.
          Lower panel: Force acting on the $C$ atoms. The squares are the calculated values,
          the solid line is obtained by derivative of the analytic curve shown with a solid
          line in the upper panel.
        }
\protect\label{figfor}
\end{figure}

As an example, we report the results of calculations for a $C_2$ dimer
(those for other molecules are very similar). The upper pannel of fig. (\ref{figfor}) 
displays the total energy Eq.\ (\ref{energy}) of the $C_2$ dimer at 5 different bond 
length (squares) along with a fourth order polynomial fit (solid line). From the fit
we find the equilibrium distance $d_0=1.26\, {\rm \AA}$ and 
the vibrational frequency $\omega _v=1860\,{\rm cm}^{-1}$ 
to be compared with the experimental values $d_0=1.24 \,{\rm \AA}$ and
$\omega _v=1854\, {\rm cm}^{-1}$.
In the lower panel of Fig. (\ref{figfor}), we compare the forces, calculated according
to Hellman-Feynman's theorem (squares) to the analytical derivative (solid line) of
the polynomial curve that interpolates the total energy values. The quality of the
force computation can be judged from the perfect matching of the points with the 
solid line.

For the systems we considered, we found the speed of convergence of the FMG to the electronic 
ground-state, for a given ionic configuration, substantially increased with respect to that 
of state-of-the-art Car-Parrinello (CP) like methods
\cite{car}, i.e. much shorter CPU times are required 
to achieve the same accuracy in the self-consistent charge density, 
total energy and ionic forces. 
For the purpose of comparison we ran a CP code based on a Damped Molecular Dynamics relaxation 
of the electronic degrees of freedom, with the maximum times step allowed to ensure stability 
during the minimization. More efficient CP schemes exist \cite{tassone}, where the convergence 
of the total energy and forces is accelerated by preconditioning techniques, which allow larger 
integration steps than those usually required in standard CP codes. Even in the latter case our 
FMG code shows better performances in minimizing the electronic energy at fixed 
atomic positions, at least for isolated molecules or clusters. Furthermore, as shown in the 
following, our approach allows to use larger time steps in Molecular Dynamics, and thus it is 
preferable for Molecular Dynamics simulations.
\begin{figure}[htb]
\centerline{\psfig{file=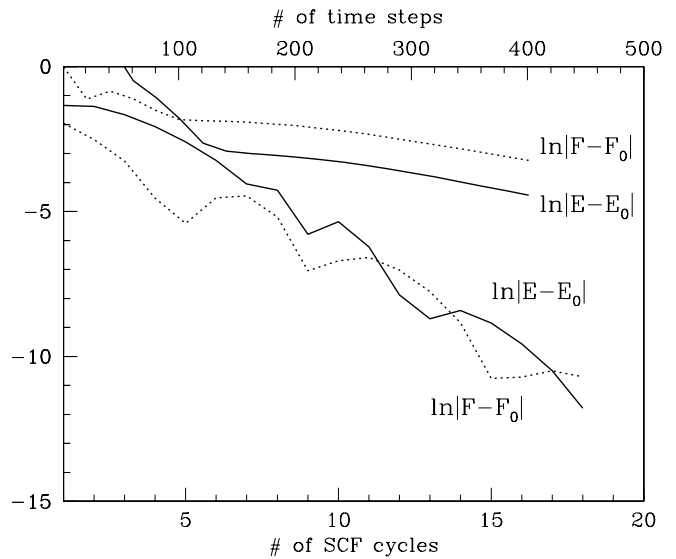, width=9cm, angle=0}}
\caption{ Comparison between the 
          convergence rates of the total energy and forces
          for the $C_2$ molecule for our FMG calculations (lower solid and dotted curves
          respectively) and for plane-wave calculations (upper solid and dotted curves).
        }
\protect\label{convrat}
\end{figure}

An example of the convergence rate of the total energy and forces is reported in 
Fig. (\ref{convrat}) for the case of a $C_2$ dimer. The logarithm of the deviation 
of the total energy $E$ and force $F$ from their values at convergence, $E_0$ and $F_0$,
are shown as a function of the number of self-consistency cycles (lower horizontal axis)
for our FMG (lower two curves) and as a function of the number of time steps 
(upper horizontal axis) for the CP run (upper two curves). Both computations required 
similar CPU time.

We did not attempt a direct comparison between our FMG and
the MG approach chosen by Briggs et al.\cite{briggs},
although we expect similar performances as far as the relaxation rate
is concerned.
We note however that,
since our FMG calculations are started on the coarsest grid,
the use of a very rough initial guess
for the electronic wavefunctions (random numbers) and density
(superposition of atomic densities) is usually sufficient
to start the computation. More educated guesses do not help in
reducing the total CPU time required to achieve convergence.
On the other hand, when we use a MG schedule, i.e. start on the
finest grid, a very good initial guess for the electron wavefunctions
is mandatory to achieve self-consistency with the same CPU time.

\begin{figure}[htb]
\centerline{\psfig{file=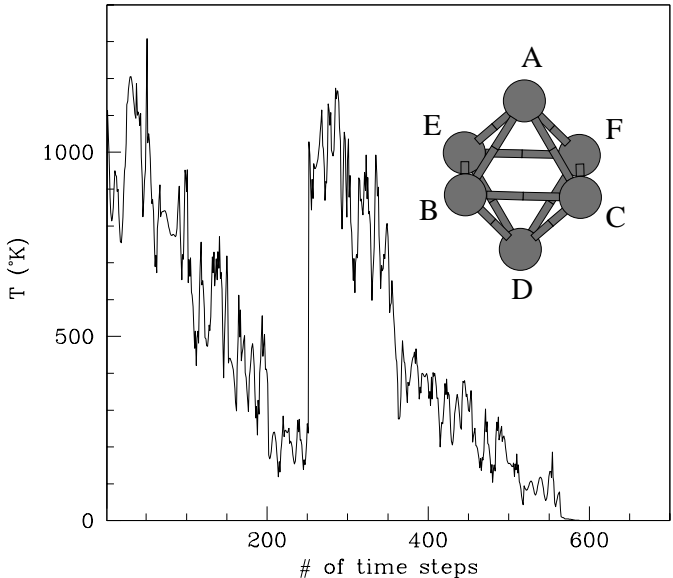, width=9cm, angle=0}}
\caption{ Simulated annealing cycle for the $Al_6$ cluster. In the last 30 steps of the run 
          a Steepest Descent relaxation on the ionic coordinates of the cluster has been 
          performed. The final structure is shown in the inset. Bonds $AE$, $AF$, $BE$, $CF$, 
          $BD$ and  $CD$ in the lowest energy structure have length $l_1=2.58\,{\rm \AA}$, 
          while 
          bonds $AB$, $AC$, $BC$, $EF$, $DE$ and $DF$ have length $l_2=2.98\,{\rm \AA}$.
        }
\protect\label{figsa}
\end{figure}

Having established the accuracy of our method for the calculation of energy and forces,
we checked its performances in the search of the lowest energy structure of a small 
metallic cluster, $Al_6$. We use a nonlocal pseudopotential, with $s$-non locality. The 
side of our cell is $L=25\,{\rm a.u.}$. It is sampled with three different grids of $13^3$, 
$25^3$ and $49^3$ points.
The effective energy cut-off on the latter grid is $9\,Ry$, which is sufficient to represent 
the pseudo charge density of $Al$ atoms in a condensed phase \cite{jones}.
Our starting geometry is arbitrary, i.e. a planar hexagonal ring of $Al$ atoms. 
To search for the lowest energy structure of this cluster, we used a Simulated Annealing 
schedule where the temperature is initially raised to $T\sim 1000\,^\circ K$ by rescaling 
the atomic velocities and then slowly reduced to $T=0$. The history of the MD run is 
recorded in Fig. (\ref{figsa}). The cluster temperature is shown as a function of the 
number of MD time steps. The time step is $\Delta t=200\,{\rm a.u.}$ and the total 
simulation time is $\sim 3$ps. After a first annealing cycle (first $\sim 200$ time steps 
in Fig. (\ref{figsa})) the cluster got caught in a local minimum of the total energy, 
corresponding to an open, quasi two-dimensional, distorted structure, $\sim 0.17\,eV/atom$ 
higher in energy than the stable structure. We thus reraised the temperature, annealed, 
and performed a Steepest Descent relaxation when in sight of the ground-state structure. 
The resulting geometry is displayed in Fig. (\ref{figsa}). This structure is characterized 
by two bond lengths, $l_1=2.58\,{\rm \AA}$ and $l_2=2.97\, {\rm \AA}$, and exhibits 
$D_{3d}$ symmetry, in agreement with previous LDA calculations \cite{jones}.

\section{The fragmentation of ${\rm Li}_{11}^+$ clusters}

After checking the accuracy and robustness of our implementation of the FMG 
scheme, we have applied it to the problem of $Li_{11}^+$ fragmentation,
which has been the subject of recent experiments \cite{broyer}. 

Experimental studies of fragmentation of small metallic clusters provide
useful information for understanding the properties of matter at the
small aggregation limit and the size-dependent evolution towards bulk behavior.
In particular, the dynamics of unimolecular clusters dissociation characterizes 
the nature of the energy partitioning among the internal vibration modes and its
understanding may provide important information on the clusters properties. 
Statistical models of energy partitioning allow, for instance, the determination 
of the binding energy of the clusters from their dissociation rates.
Cluster dissociation processes may also give some insight into the important
problem of estimating the ``melting'' temperature of the cluster, i.e.
the temperature above which the cluster becomes ``liquid'' i.e. with no well 
defined structure. Experimentally, the temperature dependence of the heat capacity 
has been measured, for instance, from the photofragmentation mass spectrum of 
alkali clusters allowing the identification of the cluster melting point \cite{schmidt}.

In most cases, neutral and singly charged metal clusters are intrinsically stable
and their dissociation is endothermic. The excess energy necessary to promote 
dissociation is provided by the photon energy in photoabsorption experiments.

Evaporation of ionized lithium clusters
has been studied recently in experiments where high-power lasers were used
to  ionize and excite clusters, and where
the evaporation into smaller products took place
in a time long with respect to the timescale set by the vibrational frequencies
of the cluster.  
A competition between the evaporation of
monomers and dimers, with the production of $Li^+_{10}$ and $Li^+_{9}$
cations respectively, has been observed \cite{broyer}.  
The dynamic behavior was found to depend critically on the ionization  
conditions.
The evaporation of dimers was found to be the
most significant fragmentation channel under irradiation with a  
$4 eV$ laser, i.e. slightly above the photoionization threshold,  
whereas suddenly both evaporative paths appeared  
as soon as the energy exceeded $4.3 eV$ and found to be of comparable
intensity when the cluster was ionized with a $4.5 eV$ laser.

If one assumes that during the photodissociation the internal energy due to multi-photon 
absorption is randomly distributed in the metastable cluster over the $3n-6$ internal modes, 
then the dissociation of a $Li_{n}^+$ cluster occurs as soon as enough internal energy 
becomes localized in a single mode, so as to overcome the fragment binding energy $D^+_n$:

\beq
D^+_n(1)=E(Li^+_{n-1})+E(Li)-E(Li^+_{n})
\protect\label{bind1}
\eeq
or
\beq
D^+_n(2)=E(Li^+_{n-2})+E(Li_2)-E(Li^+_{n})
\protect\label{bind2}
\eeq
for the two possible fragmentation channels, respectively.
Higher dissociation rates are expected for the channel characterized by a lower dissociation 
energy. From the analysis of the experimental results \cite{broyer}, however, 
non-statistical effects seem to play a major role in the process of fragmentation
of $Li^+_{11}$, i.e., even a long time after the photoexcitation
process, the internal energy appears not to be randomly distributed
among all the vibrational modes (similar non-statistical effects have been observed 
for $Na^+_4$ clusters \cite{bewig}).
As a consequence, dissociation may occur along less energetically favorable channels.
In fact, as we will show in the following, our total energy calculations
predict slightly lower dissociation energy for the dissociation channel
$Li^+_{11} \rightarrow Li^-_{10}+Li$, with the production of monomers.

Configuration Interaction (CI) calculations of the geometry and electronic structure 
of small neutral $Li_{n}$ and cationic $Li_n^+$ clusters,
with $n$ up to 10, have been reported in the literature \cite{boustani}.
The binding energies $D^+_n(1)$, 
characterizing the dissociation channel
$Li_n^+\rightarrow Li_{n-1}^++Li$,
exhibit alternating maxima and minima
for clusters with odd or even number of atoms respectively,
showing their smaller or larger stability for this dissociation channel.
For instance, smaller stability characterizes the 
dissociation of $Li_4^+$, $Li_8^+$,$Li_{10}^+$ 
with the production of monomers.
A less characteristic behavior is found in $D^+_n(2)$ for the 
$Li_n^+\rightarrow Li_{n-2}^++Li_2$ dissociation process.

Whereas, on the basis of CI calculations \cite{boustani},
small clusters (with 3 to 6 Li atoms) can be considered
as deformed sections of the (111) plane in the fcc crystals,
the structure of larger clusters can be described (although with a few 
exceptions) as more open structures, composed of deformed tetrahedrons 
appropriately
sharing their triangular sides.

The experimental binding energies for $Li_{n}$ cations were derived from the 
interpretation of photodissociation experiments \cite{brechi}.
They are determined from a statistical treatment of unimolecular decay rates, where 
the energy is assumed to be statistically distributed over the various nuclear degrees
of freedom. In Ref. \cite{brechi} the binding energies of the cationic $Li^+_{n}$ 
clusters with $n$ up to 10 obtained by this phenomenological method are compared with 
results of CI calculations. The discrepancy between CI and statistical cluster 
dissociation energies can however be as large as $0.4$ eV.

The same statistical model predicts \cite{brechi} the values 
$D_{11}^+(1)=1.30\,eV$ and $D_{11}^+(2)=1.07\,eV$
for the two evaporative channels of $Li^+_{11}$ clusters,
but no first principles calculations for this cluster exist
to support these estimates.
We provide here results, obtained with our FMG method, for the
dissociation energies of $Li^+_{11}$ clusters along their two main 
fragmentation paths. At variance with the results \cite{brechi} quoted above,
we find that the fragmentation into monomers is slightly 
favored on energetics grounds, i.e. $D_{11}(1)^+$ is lower than $D_{11}(2)^+$. 

The Full Multigrid scheme is implemented on a set of three grids of 
$N^3$ points, with $N=21,41,81$.
We used the LSD version of Eq. (\ref{energy}),
as described in Section 2. The electron-ion interaction is modeled
with a nonlocal pseudopotential\cite{bachelet}, with $s$-non locality.
The side of the cell is $L=36\,{\rm a.u.}$. This means that on the finest grid
the effective energy cut-off is $12\,Ry$, which is sufficient 
to represent the pseudo charge density 
of $Li$ atoms in a condensed phase.
The values $d_{eq}=2.65\,{\rm \AA}$ ($2.67$) and $\omega _v =348\,{\rm cm}^{-1}$ ($351$),
obtained for the equilibrium distance and the vibrational frequency 
(in parenthesis we report the experimental values) of the $Li_2$ dimer
show the reliability of the pseudopotential 
to provide an accurate description of the system.

An unbiased search, Simulated Annealing, is used to obtain the lowest energy
atomic configurations for the cations $Li_n^+$, with $n=9$,$10$ and $11$.
Annealing is typically started from a randomly generated initial configuration
for the ionic cores; the ``temperature'' (defined in terms of the 
total ionic kinetic energy) is initially raised to 
$T\sim 800-1000\,^\circ K$ by rescaling the atomic velocities and then slowly reduced to $T=0$. 
We used a simulation time step $\Delta t=200\,{\rm a.u.}$.

\begin{figure}[htb]
\centerline{\psfig{file=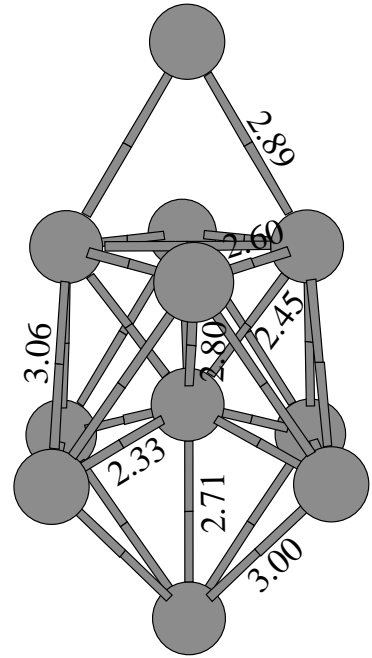, width=5cm, angle=0}}
\caption{
Equilibrium geometry of the $Li_{11}^+$ 
cation. Bond lengths are shown in \AA.
}
\protect\label{figeq1}
\end{figure}

The resulting structure for $Li^+_{11}$, very similar to the
lowest energy structure of the neutral $Li_{11}$ cluster, 
is shown in Fig. (\ref{figeq1}). The cluster has $C_{2v}$ symmetry and 
the relevant bond lengths are reported in the Figure
(for comparison, we remind that the experimental interatomic distance 
in the fcc Li crystal is $3.10\,{\rm \AA}$).
We also calculate the dipole polarizability of the cluster by
applying a small uniform electric field ${\bf E}$ and by computing 
the resulting dipole moment ${\bf P}$. The field intensity must be small 
enough to produce a dipole moment proportional to the field. 
The polarizability is then evaluated as $\alpha=P/E$. We find 
$\alpha_{\parallel}=128\,{\rm \AA}^3$ and $\alpha_{\perp}=76\,{\rm \AA}^3$ 
for the polarizabilities along the cluster axis connecting the two cap atoms 
at the far ends of the molecule, and perpendicular to it.
\begin{figure}[htb]
\centerline{\psfig{file=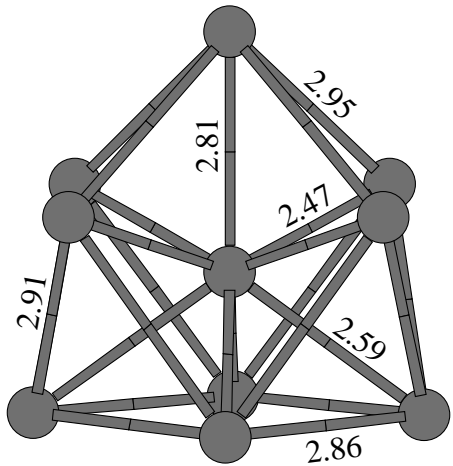, width=6cm, angle=0}}
\caption{
Equilibrium geometry of the $Li_{10}^+$ 
cation. 
}
\protect\label{figeq2}
\end{figure}

\begin{figure}[htb]
\centerline{\psfig{file=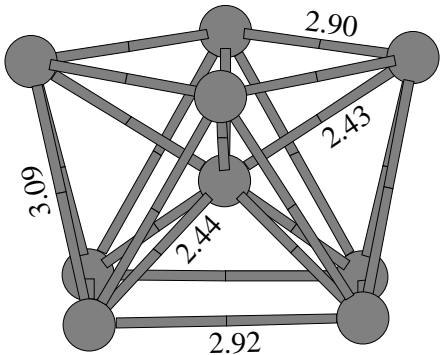, width=6cm, angle=0}}
\caption{
Equilibrium geometry of the $Li_{9}^+$ 
cation. 
}
\protect\label{figeq3}
\end{figure}
The lowest energy structures of $Li^+_{10}$ and $Li^+_{9}$, i.e. the
fragments corresponding to the relevant dissociation paths
of $Li_{11}^+$, are shown in Fig.(\ref{figeq2}) and Fig.(\ref{figeq3}), respectively. 
Both structures have $C_{4v}$ symmetry.
By evaluating the dissociation energies 
for the two channels, 
according to the definitions (\ref{bind1}) and (\ref{bind2}),
we find that the dissociation with the production 
of a monomer is energetically slightly favored over that producing a dimer.
In fact, our calculated values for these dissociation energies are 
$D_{11}^+(1)=1.07 eV$ and $D_{11}^+(2)=1.11 eV$.

These findings seem to support the claim, as suggested in Ref.\cite{broyer}, that the 
fragmentation process is dominated by non-statistical effects, which 
may favor dissociation along the less energetically favored channels.

The sophisticated calculations of the dynamics of highly excited cluster which are 
required to explain in detail the experimental results of Ref.\cite{broyer} are of course 
beyond the applicability of ground-state theories such as DFT.  
High level ab-initio calculations of the excited states incorporating large
scale valence electron configuration interaction are probably necessary to
calculate reliably the optical response of the clusters and thus to predict the
dissociation pathways.  
We believe, however, that informations about
the structure and energetics such as those reported here may provide a useful
input for more refined calculations where a realistic description of the  
excitation process and the way in which the energy is redistributed among
the excited states are treated.  

\section{Summary and Conclusions}

We presented a method to perform ab-initio electronic structure 
calculations in real-space by 
using a local and accurate discretization of the electron-ion 
Hamiltonian within DFT.
We use optimized pseudopotentials particularly suited for calculations in
real-space. The force acting on the ions are calculated accurately and the
feasibility of ab-initio Molecular Dynamics is demonstrated.  At variance with
a recently proposed Mehrstellen-MG scheme, our method implements a
FMG schedule that uses a set of coarser grid where the KS equations are solved 
approximately. We found that, for cluster calculations, our code has performances 
comparable or superior  to the most efficient PW codes available nowadays. 

The method has been applied to ab-initio MD simulations with large time steps 
without loss of accuracy in the total energy during the simulations.  As for 
many real-space schemes proposed recently, one of the major advantages of this
FMG scheme is that it can be readily adapted to run efficiently on parallel computer
architectures: all operations, except for the orthogonalization of the KS orbitals,
are short-ranged. For the simple molecules and clusters investigated here,
the largest fraction (80- 90$\%$) of the CPU time is spent in performing relaxations, 
which execute very efficiently on computers with parallel architecture. 
We have studied with this method the energetics of fragmentation of $Li^+_{11}$ 
clusters and confirmed the importance of the statistical effects in their
dissociation.

\acknowledgments
We acknowledge useful discussions with 
E.L.Briggs, F.Gygi and G.Galli.
One of us (P.B.) acknowledges financial support from the Fonds National Suisse pour
la recherche Scientifique and useful discussions with U.Landman, R.Barnett and S.Wei.


\bigskip

\begin{thebibliography}{99}

\bibitem{cheli}
{ J.R.Chelikowsky, N.Troullier and Y.Saad, Phys. Rev. Lett. {\bf 72}, 1240
  (1994);J.R.Chelikowsky, N.Troullier, K.Wu and Y.Saad, Phys. Rev. B {\bf 50},
  11355 (1994)}.

\bibitem{galli}
{ F.Gygi and G.Galli, Phys. Rev. B {\bf 52}, R2229 (1995)}.

\bibitem{gygi}
{ F.Gygi, Europhys. Lett. {\bf 19}, 6617 (1992); F.Gygi, Phys. Rev. B {\bf 48},
  11692 (1993)}.

\bibitem{zumbach}
{ G.Zumbach, N.A.Modine and E.Kaxiras, Sol. State Commun. {\bf 99}, 57 (1996);
  N.A.Modine, G.Zumbach and E.Kaxiras, Phys. Rev. B{\bf 55}, 10289 (1997)}.

\bibitem{cho}
{ K.Cho, T.A.Arias, J.D.Joannopoulos and P.K.Lam, Phys. Rev. Lett. {\bf 71},
  1808 (1993)}.

\bibitem{wei}
{ S.Wei and M.Y.Chou, Phys. Rev. Lett. {\bf 76}, (1995)}.

\bibitem{white}
{ S.R.White, J.W.Wilkins and M.P.Teter, Phys. Rev. B {\bf 39}, 5819 (1989)}.

\bibitem{tsu}
{ E.Tsuchida and M. Tsukada, Phys. Rev. B {\bf 52}, 5573, (1995)}.

\bibitem{iyer}
{ K.A.Iyer, M.P.Merrick and T.L.Beck, J. Chem. Phys. {\bf 103}, 227 (1995)}.

\bibitem{briggs}
{ E.L.Briggs, D.J. Sullivan and J.Bernholc, Phys. Rev. B {\bf 52}, R5471
  (1995)}.

\bibitem{brig96}
{ E.L.Briggs, D.J. Sullivan and J.Bernholc, Phys. Rev. B {\bf 54}, 14362
  (1996)}.

\bibitem{mg}
{ A.Brandt, Math. Comput. {\bf 31}, 333 (1977); GDM Studien, {\bf 85}, 1
  (1984)}.

\bibitem{shaw}
{ {\it Relaxation Methods}, F.S. Shaw ed., Dover Publications Inc. (1953)}.

\bibitem{hackbush}
{ W.Hackbush, {\it Multigrid Methods and Applications}, Springer-Verlag, Berlin
  (1985)}.

\bibitem{recipes}
{ Numerical recipies in FORTRAN, The Art of Scientific Computing, Second
  Edition, William H. Press, Saul A. Teukolsky, William T. Vetterling, Brian P.
  Flannery, Cambridge University Press }.

\bibitem{costiner}
{ S.Costiner and S.Ta'asan, Phys. Rev. E {\bf 51}, 3704 (1995)}.

\bibitem{ks}
{W.Kohn and L.J.Sham, Phys. Rev. {\bf 140} A1133 (1965)}.

\bibitem{hk}
{ P.Hohenberg and W.Kohn, Phys. Rev. {\bf 136}, B864 (1964);}.

\bibitem{kleinman}
{ L.Kleinman and D.M.Bylander, Phys. Rev. Lett. {\bf 48}, 1425 (1982)}.

\bibitem{martins}
{ N.Troullier and J.L.Martins, Phys. Rev. B {\bf 43}, 8861 (1991)}.

\bibitem{vosko}
{ U.von Barth and L.Hedin, J. Phys. C {\bf 5}, 1629 (1972); S.H.Vosko et al.
  Can. J. Phys. {\bf 58}, 1200 (1980)}.

\bibitem{Perdew}
{ J.P. Perdew and A. Zunger, Phys. Rev. B{\bf 23}, 5048 (1981)}.

\bibitem{Ceperley}
{ D.M. Ceperley and B.J. Alder, Phys. Rev. Lett. {\bf 45}, 566 (1980)}.

\bibitem{mehr}
{ L.Collatz, {\it The numerical Treatment of Differential Equations},
  Spinger-Verlag, Berlin, 1960}.

\bibitem{king}
{ R.D.King-Smith, M.C.Payne and J.S.Lin, Phys. Rev. B{\bf 44}, 13063 (1991)}.

\bibitem{bernholc}
{ J.Bernholc, J.-Y. Yi and D.J.Sullivan, Faraday Disc. Chem. Soc. {\bf 92}, 217
  (1991)}.

\bibitem{bachelet}
{ G.B.Bachelet, D.R. Hamann and M.Schluter, Phys. Rev. B {\bf 26}, 4199
  (1982)}.

\bibitem{car}
{R.Car and M.Parrinello, Phys. Rev. Lett. {\bf 55}, 2471 (1985)}.

\bibitem{tassone}
{F.Tassone, F.Mauri and R.Car, Phys. Rev. B {\bf 50}, 10561 (1994)}.

\bibitem{jones}
{ R.O.Jones, Phys. Rev. Lett. {\bf 67}, 224 (1991)}.

\bibitem{broyer}
{ R.Antoine, D.Rayane, Ph.Dugourd,B.Vezin,B.Tribollet and M.Broyer,Surface
  Review and Letters, {\bf 3}, 545 (1996)}.

\bibitem{schmidt}
{M.Schmidt,R.Kusche,W.Kronmuller,B.Issendorff and H.Haberland, Phys. Rev. Lett.
  {\bf 79}, 99 (1997)}.

\bibitem{bewig}
{L. Bewig, U. Buck, Ch. Mehlmann and M. Winter, J. Chem. Phys. {\bf 100}, 2765
  (1994)}.

\bibitem{boustani}
{I.Boustani, W.Pewerstorf, P.Fantucci, V.Bonacic-Koutecky and J.Koutecky, Phys.
  Rev. B{\bf 35}, 9437 (1987)}.

\bibitem{brechi}
{C.Brechignac, H.Busch, Ph.Cahuzac and J.Leygnier, J.Chem.Phys. {\bf 101}, 6992
  (1994)}.

\end{thebibliography}
\end{document}